\documentclass{moriond}

 \usepackage{graphicx,amssymb,amsmath,stmaryrd,makecell}

\usepackage[square,sort,comma,numbers]{natbib}

\def\be{\begin{equation}}
\def\ee{\end{equation}}
\def\bea{\begin{eqnarray}}
\def\eea{\end{eqnarray}}

\newcommand{\Photo}{}

\begin{document}
\vspace*{4cm}
\title{Cosmological constraints from the thermal Sunyaev Zeldovich power spectrum?}

\author{Boris Bolliet}

\address{Jodrell Bank Centre for Astrophysics, School of Physics and Astronomy,
The University of Manchester, Manchester, M13 9PL, U.K.}

\maketitle\abstracts{
The latest Planck results on the power spectrum of the Compton-$y$ parameter are the most accurate probe of the thermal Sunyaev Zeldovich (tSZ) effect caused by resolved and un-resolved clusters of galaxies. On large angular scales, the power spectrum amplitude is mostly due to the statistical distribution of clusters in the sky and therefore is an indirect probe of the clustering of matter, $\sigma_8$, as well as the angular diameter distance that depends on other cosmological parameters such as the Hubble parameter, $h$, and the matter density $\Omega_m$. Here I discuss the constraining power of the tSZ  power spectrum, in light of the Planck Compton-$y$ map data, and our current understanding and measurements of the intra-cluster medium.}

The Planck mission is a benchmark for precision cosmology. Three major outputs deserve a special emphasis: first, the Planck Collaboration enabled us to measure the cosmological parameters of the $\Lambda$CDM model with unprecedented accuracy, significantly improving over WMAP results. This was crucial for ruling out some models of inflation and to pave the way for the forthcoming CMB primordial gravitational wave experiments. Second, the Planck Collaboration produced the first ever full-sky map of the lensing potential. This characterises the distribution of the dark matter over the entire observable universe. Third, it measured the locations, sizes and SZ fluxes of more than four hundred clusters of galaxies as well as the diffuse SZ effect coming from \textit{all} the clusters of galaxies in the observable universe: the full-sky Compton-$y$ map \cite{Aghanim:2015eva}. This last point is the focus of this proceeding. 

The thermal SZ effect is the Compton scattering of CMB photons by the hot gas of electrons that surrounds galaxy clusters, filling the potential wells created by the dark matter \cite{1972CoASP...4..173S}. It is a frequency dependent effect which results in a decrease of the intensity of the CMB spectrum at low frequencies and an increase at high frequencies, with a crossover frequency at 217GHz.  The SZ effect is quantified by the Compton-$y$ parameter, proportional to the integral of the electron temperature over the intra-cluster medium (ICM). For the tSZ effect,  the Compton-$y$ parameter is a positive dimensionless number of order $10^{-6}$. Positive because the electron temperature in the dark matter halo is higher than the CMB temperature: energy flows from the electrons to the CMB photons. 

Thanks to the unique frequency signature of the SZ effect, the Planck Collaboration was able to extract the full-sky Compton-$y$ parameter map using components separation algorithms such as NILC on its temperature maps \cite{2011MNRAS.410.2481R}. 
As far as cosmology is concerned, rather than the astrophysics of the ICM, two types of analyses can be carried out using the SZ data. On the one hand, the \textit{number count} analysis exploits the statistical distribution of the resolved clusters with respect to their masses and redshifts \cite{Ade:2015fva}. On the other hand, the \textit{power spectrum} analysis exploits the two-point correlation function of the full-sky Compton-$y$ map, to which all clusters (resolved and uresolved) are contributing \cite{Aghanim:2015eva}. Here we are interested in the latter. 
\\

The component separation algorithms can not produce completely clean individual component maps, there is always a leakage of the other components into the individual component maps. In particular, the Compton-$y$ map is significantly contaminated by the cosmic infrared background (CIB), the emission from radio point sources (RS), infra-red point sources (IR) and instrumental correlated noise (CN). Hence, a realistic decomposition of the angular power spectrum of the Compton-$y$ parameter, $C_{\ell}^{y^{2}}$, computed from the map is
\begin{equation}
C_{\ell}^{y^{2}}  =C_{\ell}^{{\mathrm{tSZ}}}+A_{{\mathrm{CIB}}}\hat{C}_{\ell}^{{\mathrm{CIB}}}+A_{{\mathrm{IR}}}\hat{C}_{\ell}^{{\mathrm{IR}}}+A_{{\mathrm{RS}}}\hat{C}_{\ell}^{{\mathrm{RS}}}+A_{{\mathrm{CN}}}\hat{C}_{\ell}^{{\mathrm{CN}}},\label{eq:model-2}
\end{equation}
where $C_{\ell}^{{\mathrm{tSZ}}}$ represents the contribution from the tSZ effect, while the other terms are the contributions from the other sources and noise. These are tabulated templates and their amplitudes are treated as nuisance parameters. These contaminants are dominant on small scales, typically $\ell\gtrsim 2000$, while the tSZ effect represents the main contribution on larger scales.  The noise amplitude, $A_{{\mathrm{CN}}}$, is fixed by comparing the power on the largest multipoles with the noise template, because the signal is largely dominated by noise on these scales. See \cite{Aghanim:2015eva} for the original Planck analysis. 

What makes the angular power spectrum of the tSZ effect a probe of cosmology? Let us address this question by looking at the formula for the tSZ power spectrum  \cite{Komatsu:1999ev}: 
\begin{equation}
C_{\ell}^{{\mathrm{tSZ}}}= \int\mathrm{d}z\frac{\mathrm{d}V}{\mathrm{d}z\mathrm{d}\Omega}\int\mathrm{d}M\frac{\mathrm{d}n}{\mathrm{d}M}\left|y_{\ell}\left(M,z\right)\right|^{2}.\label{eq:cltSZ-1}
\end{equation}
This is a beautiful and simple formula. The first integral, over redshift, with the differential volume element, means that the tSZ power spectrum contains the contribution from galaxy clusters of \textit{all ages}, all the way to the time when they started forming. The second integral, over the masses, with the halo mass function $\mathrm{d}n/\mathrm{d} M$, means that the tSZ power spectrum contains the contributions from galaxy clusters of \textit{all masses}. The last term is the squared amplitude of two dimensional Fourier transform of the electron temperature (or pressure) profile. This is where the SZ effect is hidden, and the amazing feature is the redshift dependency. The term $|y_\ell |^2$ has a redshift dependence, because the angular sizes of a cluster on the map depends on how far it is from us: an effect of perspective. But this is the \textit{only}  redshift dependence, in particular, the \textit{amplitude} of the SZ effect is redshift independent.

\begin{figure}
\begin{centering}
\includegraphics[height=5cm]{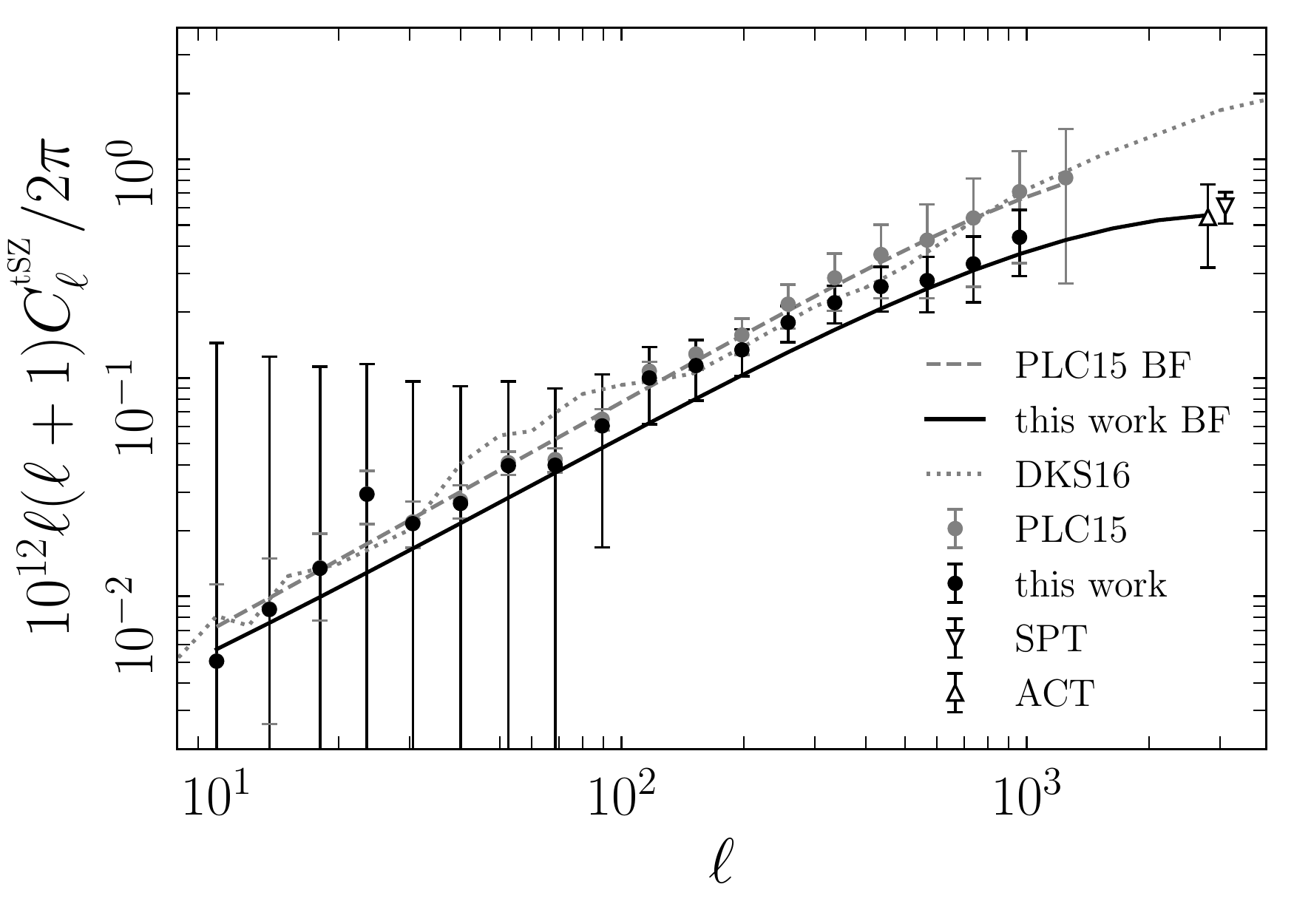}
\par\end{centering}
\caption[]{The black line is the best-fitting tSZ power spectrum from our analysis. The dashed grey line is that of the Planck 2015 analysis. The black filled circles show the data points and error bar with the foreground spectra marginalised over. The grey circles and error bars come from the Planck 2015 analysis. We also show the ACT and SPT data points for
 comparison, as well as the simulation results of \protect\cite{Dolag:2015dta} (DKS16, grey dotted line).}
\label{fig:radish}
\end{figure}

At a fixed redshift, clusters are randomly distributed in space. The power spectrum of randomly distributed point sources is a Poisson spectrum, i.e., $C_\ell \simeq \mathrm{constant}$. Modelling clusters as point sources is correct as long as one studies the two-point correlation function on scales larger than the spatial extension of the cluster, typically for multipoles $\ell\lesssim 10^3$.   At these multipoles the tSZ power spectrum is therefore a probe of the statistical distribution of clusters on cosmological scales, and is not sensitive to the physical processes at play in the ICM. On scales smaller than the spatial extension of clusters the power spectrum decreases, and depends on the hydrodynamical states of the ICM. The tSZ power spectrum is generally presented in units $D_\ell = \ell(\ell+1)C_\ell/2\pi$ and peaks at around $\ell=3000$, see figure \ref{fig:radish}. 

The halo mass function quantifies the number of clusters per unit mass, per unit volume, at a given redshift. In practice, it is  an analytical formula inspired by the Press-Schechter formalism that is fitted to the results of N-body simulations \cite{Jenkins:2000bv}. Its main variable is the variance of the over-density field smoothed over spherical regions that enclose the mass $M$ (the halo mass): $\mathrm{d}n/\mathrm{d}M \propto f(\sigma, z)$ with
\begin{equation}
\sigma^{2}\left(M,z\right)\equiv\int\frac{\mathrm{d}k}{k}\frac{k^{3}}{2\pi^{2}}P\left(k,z\right)W^2\left(kR\right)\label{eq:sigma2m}\,,
\end{equation} 
where $W$ is the top-hap window function that defines the spherical regions with size $R=[3M/4\pi\rho_{m0}]^{1/3}$ and $P(k,z)$ is the linear matter power spectrum. This is why the tSZ power spectrum at large scales exhibits a strong dependence on the amplitude of clustering of matter, i.e., $\sigma_8$. 

The electron pressure profile, that enters $|y_\ell |^2$, is crucial for both the small and large scale tSZ power. On small scales, the morphology of the halos determines the spatial variation of the electron pressure with respect to the centre of the halo which in turns affects the tSZ power. On large scales, the total integrated pressure contributes to the amplitude of the tSZ power spectrum, this essentially depends on the mass of the clusters rather than their morphology.

The parameterisation of the electron pressure profile, as a function of the cluster mass relies on the scaling relation between the SZ flux and the mass inferred from X-ray observations. However, when one compares the mass inferred from X-ray observations to their `true' mass, obtained for instance from lensing, there is an important mismatch which remains partly unexplained. It leads to the introduction of a bias parameter $B$ in the formula of the electron pressure profile, the so-called \textit{mass bias}. 


We implemented the computation of the tSZ power spectrum  in \verb|CLASS| and studied the scaling of the tSZ power with the cosmological parameters and the mass bias \cite{Bolliet:2017lha}. On large scales the dependence is well approximated by 
\begin{equation}
\ell(\ell+1)C_{\ell}^{{\mathrm{tSZ}}}\propto\sigma_{{{\mathrm{8}}}}^{8.1}\Omega_{{\mathrm{m}}}^{3.2}B^{-3.2}h^{-1.7}\,\,\,\,\mathrm{for}\,\,\,\,\,\mathrm{\ell}\lesssim10^{3}.\label{eq:scaling}
\end{equation}
Then, with \verb|Montepython| \cite{Brinckmann:2018cvx}, we searched for the cosmological parameters and foreground amplitudes that minimise the log likelihood $-2\ln\mathcal{L}=\chi^{2}+\ln|M|+{\rm const.}$, where 
\begin{equation}
\chi^{2}\equiv\sum_{a\leq a^{\prime}}\left(C_{\ell_{{\mathrm{eff}}}^{a}}^{y^{2}}-\hat{C}_{\ell_{{\mathrm{eff}}}^{a}}^{y^{2}}\right)\left[M^{-1}\right]_{aa^{\prime}}\left(C_{\ell_{{\mathrm{eff}}}^{a^{\prime}}}^{y^{2}}-\hat{C}_{\ell_{{\mathrm{eff}}}^{a^{\prime}}}^{y^{2}}\right),\label{eq:likelihood}
\end{equation}
where $a,a^{\prime}$ are indices for the multipole bins running from
$a=1$ to $a=18$, $C_{\ell_{{\mathrm{eff}}}^{a}}^{y^{2}}$ is computed according to Eq.~\eqref{eq:model-2} and \eqref{eq:cltSZ-1}, $\hat{C}_{\ell_{{\mathrm{eff}}}^{a}}^{y^{2}}$ is computed from the Planck Compton-$y$ map \cite{Aghanim:2015eva}, and $M_{aa^{\prime}}$ is the covariance matrix. The covariance matrix has two main parts, the Gaussian part which comes from the cosmic sampling variance and Gaussian instrumental noise, and a non-Gaussian part which comes from the fact that the halo distribution is not Gaussian. The non-Gaussian covariance can be computed from the four-point correlation function of the Compton-$y$, or analytically using the trispectrum formula following   \cite{Komatsu:2002wc}. We adopted the later strategy.  

The non-Gaussian covariance was omitted in the Planck analysis, but it turns out that it is an essential piece for the cosmological parameter extraction. It dominates over the Gaussian covariance on large scales: this is because the low redshift halos may cover a significant fraction of the sky and lead to non-negligible multipole-to-multipole correlation. At the end, it appears that the non-Gaussian part of the covariance drives the error bars on the cosmological parameters.  A way around, could be to identify and apply a mask to the halos that cause these correlation. 

In our analysis, we varied the six base cosmological parameters, the three foreground amplitudes, and the mass bias. We computed the trispectrum at each step of the MCMC, as it depends on the cosmological parameters and mass bias. Moreover, we used the information contained in the Planck SZ catalogue of clusters to impose an upper bound on the combined foregrounds. Indeed the projection of the SZ fluxes coming from the Planck catalogues on a sky map yields a lower bound for the tSZ power spectrum. Several authors have already revisited and extended the Planck analysis, including \cite{Horowitz:2016dwk,Salvati:2017rsn,Hurier:2017jgi}, nevertheless this is the first time that the analysis is carried out consistently, with all the relevant pieces together. 
We found a constraint on the parameter combination $F\equiv\sigma_{{{\mathrm{8}}}}\left(\Omega_{{\mathrm{m}}}/B\right)^{0.40}h^{-0.21}$, motivated by the scaling shown in Eq.~\eqref{eq:scaling}. Taking into account the trispectrum resulted in an increase of the error bar on $F$ by more than a factor of two. We measured $F$ with a 2.5\% accuracy:  $F=0.460\pm0.012$ at 68\%CL. Then, we marginalized the measured Compton-$y$ map power spectrum over the foregrounds and correlated noise to deduce the tSZ power spectrum amplitudes and error bars in each multipole bins. While the error bars on the tSZ power spectrum are significantly larger than the one found in the Planck analysis, due to the trispectrum, the amplitude agree well at low multipole. Moreover, our tSZ power spectrum starts departing from the Planck one from $\ell\simeq300$ with smaller amplitudes towards higher multipole, which nicely fits the ACT and SPT data points at $\ell=3000$ (see figure \ref{fig:radish}).  

Does the SZ data agree with primordial CMB  temperature anisotropy constraints? The Planck 2015 chains `TT+lowP' give $\sigma_{{{\mathrm{8}}}}\Omega_{{\mathrm{m}}}^{0.40}h^{-0.21}=0.568\pm 0.015$  at 68\%CL. Since the SZ data constrains a combination of these parameters and the mass bias, a more appropriate way to ask the question is: what value of the mass bias makes the SZ and primordial CMB data consistent with one another. The answer is $B=1.71\pm0.17$ at 68\% CL. The SZ cluster count analysis yields a very similar mass bias \cite{Ade:2015fva}, as well as the joint SZ and 2MASS power spectrum analysis \cite{Makiya:2018pda}. These results are puzzling. Indeed, for both probes to be consistent, it means that the halo mass of the Planck clusters, inferred from X-ray measurements of the gas temperature, is more than forty percents lower than their true mass. 

Numerical simulations of the hydrodynamical processes at play in the electron gas of dark matter halos surrounding galaxy clusters, can explain a bias of about twenty percents whose source is departure from the hydrostatical equilibrium assumption that is used in the mass calibration of the Planck clusters, see e.g., \cite{2012ApJ...758...74B}.  Departure from hydrostatic equilibrium arises from non-thermal pressure caused by processes such as turbulence, magnetic fields or cosmic rays. From the observational side, a recent study of clusters from the XMM-Newton survey also pointed toward a bias of order fifteen percent due to non-thermal pressure \cite{Eckert:2018mlz}. So, there is a remaining twenty percents of bias that remains unexplained. It could still be partly associated to systematics uncertainty in the X-ray data, or it could also be a hint for new physics such as dark energy or massive neutrinos. 

The constraining power of the SZ data for cosmological models is therefore limited by an unexplained mass bias. Hence, the tSZ data is not yet competitive with other probes such as CMB lensing or galaxy power spectrum but we can be hopeful that ongoing surveys such as NIKA2 for the SZ part, and forthcoming X-ray surveys such as eROSITA will shed more light on the physics of the ICM and solve the bias puzzle within the next couple of years. Given the 2.5\% measurement of the parameter combination $F$, if the mass bias becomes well understood and accurately modelled, the SZ data will become competitive for cosmological constraints.

\section*{Acknowledgments}

I am very grateful to the organisers and the people I met at the conference, to my co-authors Barbara Comis, Eiichiro Komatsu,  Juan Macías-Pérez, and to Jens Chluba for his comments on this manuscript. My work is supported by Jens Chluba's ERC grant CMBSPEC No. 725456.


\bibliographystyle{mnras}
\bibliography{moriond}

\end{document}